\DocumentMetadata{%
	lang=en-US,
	pdfstandard	= A-3u,
	pdfversion	= 1.0,
}

\documentclass[balance,upint,varvw,hyphenate,barcolor=black,german,nocopyright]{asmejour}%

\allowdisplaybreaks 

\hypersetup{%
	pdfauthor={Tihomir Varchev},                       		   	
	pdftitle={Exergy-Anergy Representation of Turbomachine Performance Characteristics},                  	
	pdfkeywords={Performance Maps, Exergy, Anergy, Turbine, Compressor, Characteristics, Performance Synthesis},
	pdfsubject = {Introducing a novel type of turbomachine performance characteristics based on exergy and anergy},			
}

\JourName{Turbomachinery}
\PreprintString{PREPRINT}

\usepackage{graphicx}
\usepackage{subfloat}
\usepackage{multirow}
\usepackage{booktabs}
\usepackage{multirow}
\usepackage{siunitx}
\usepackage{longtable,tabularx}
\usepackage{array}

\usepackage{makecell}

\newcommand{\ptot}[1]{p_{{\mathrm{t,\,#1}}}}
\newcommand{\mdot}[1]{\dot{m}_\mathrm{#1}}

\newcommand{\ttot}[1]{T_{{\mathrm{t,\,#1}}}}
\newcommand{\htot}[1]{h_{{\mathrm{t,\,#1}}}}
\newcommand{\entr}[1]{s_{{\mathrm{#1}}}}

\newcommand{\cp}{\ensuremath{c_p}}
\newcommand{\epar}{\ensuremath{\hat{\epsilon}}}
\newcommand{\bpar}{\ensuremath{\hat{\beta}}}
\newcommand{\eisc}{\ensuremath{\eta_{\mathrm{C,\,is.}}}}
\newcommand{\eist}{\ensuremath{\eta_{\mathrm{T,\,is.}}}}
\newcommand{\prc}{\ensuremath{\Pi_{\mathrm{C}}}}
\newcommand{\prt}{\ensuremath{\Pi_{\mathrm{T}}}}
\newcommand{\kmoqk}{\ensuremath{\frac{\gamma - 1}{\gamma}}}
\newcommand{\kqkmo}{\ensuremath{\frac{\gamma}{\gamma - 1}}}

\begin{document}


\SetAuthorBlock{Tihomir Varchev\CorrespondingAuthor}{%
  Institute of Aircraft Propulsion Systems,\\
  University of Stuttgart,\\
  Stuttgart 70569, Germany\\
  email: tihomir.varchev@ila.uni-stuttgart.de
}

\SetAuthorBlock{Yiwen Yuan}{%
  Institute of Aircraft Propulsion Systems,\\
  University of Stuttgart,\\
  Stuttgart 70569, Germany\\
}

\SetAuthorBlock{Tobias Schateikis}{%
  Institute of Aircraft Propulsion Systems,\\
  University of Stuttgart,\\
  Stuttgart 70569, Germany\\
}

\SetAuthorBlock{Stephan Staudacher}{%
  Institute of Aircraft Propulsion Systems,\\
  University of Stuttgart,\\
  Stuttgart 70569, Germany\\
}


\title{Exergy-Anergy Representation of Turbomachine Performance Characteristics}


\begin{abstract}
The performance synthesis calculation of aircraft engines relies on tabulated datasets for the simulation of the complex turbomachinery components at acceptable runtimes. Typically, these so-called performance maps express the change of the fluid's energetic state over the component in terms of the total pressure ratio and the isentropic efficiency of the process. However, the definitions of both parameters are different for compressors and turbines and the isentropic efficiency is in both cases not defined for a unity pressure ratio. Moreover, this parameter combination is not directly applicable when modelling the engine at the aircraft level, which is required for modern highly integrated aircraft design. As exergy analysis is an established tool for aircraft design asessment, novel performance maps are proposed that describe the change in exergy and anergy over a given turbomachine component. Both changes are expressed as non-dimensional parameters whose definition is consistent for compressors and turbines, and is compatible with a unity pressure ratio. It is shown that the presented exergy-anergy maps are obtainable on a standard turbomachine test bed or through transformation of existing maps. It is highlighted using examples, that the conversion between existing maps and exergy-anergy maps is completely lossless. The different operating regimes of the turbomachines are clearly distinguishable in the novel map representation, which allows an assessment of the map's physical consistency. It is therefore concluded, that the exergy-anergy maps are an important alternative to the performance map variants established today.
\end{abstract}

\date{Preprint version, \today}
\date{Submitted to the ASME Journal of Turbomachinery on September 1, 2026}

\maketitle 


\section{Introduction}

Performance synthesis calculation is of significant importance during the entire life cycle of a modern aircraft engine from its preliminary design~\cite{Donus2011} up to monitoring its health state during operation~\cite{Weiss2022,Varchev2024}. The engine is represented by a so-called engine performance model which describes the engine components, the mechanical connections between them, and the engine's primary and secondary air systems~\cite{Kurzke2018}. Each engine component is represented by a deterministic sub-model which maps the gas path conditions at its inlet to the gas path conditions at its outlet, thus describing the change to the aero-thermodynamic state of the gas across the component~\cite{Walsh2004}. The performance synthesis calculation is carried out component-wise in the direction of the flow, ensuring the conservation of mass and energy along the gas path. While analytic modelling is sufficient for simpler components, e.g. ducts, the thermodynamics of the turbomachine components is captured in tabulated datasets~\cite{Nielsen2005}. These so-called performance maps are based on non-dimensional groups that capture the turbomachine's performance at acceptable computational times~\cite{Nielsen2005}. The non-dimensional performance maps are valid for geometrically similar machines at a given entry swirl angle.

\renewcommand{\arraystretch}{2}
\begin{table*}[h!]
\caption{Established Types of Compressor Performance Maps}
  \centering
  \begin{tabularx}{\linewidth}{l|l|r@{$\:$}l|X X X}
    \toprule
    \multicolumn{4}{c|}{\multirow[c]{1}{*}{Non-Dimensional Group}} & \multicolumn{3}{c}{\multirow[c]{1}{*}{Used in Maps of Type}}\\\cline{1-4}\cline{5-7}
    Name & Describes & \multicolumn{2}{c|}{Definition} & $\varphi$-$\Psi$-$\eta$ & $Q$-$\Pi$-$\eta$ & $Q$-$\Pi$-$\hat{M}$ \\
    \midrule
    Flow Coefficient & Rotor Incidence & $\varphi$&$= \frac{c_\mathrm{ax}}{u}$ & $\checkmark$ & & \\
    Work Coefficient & Shaft-Work Input & $\Psi_{\mathrm{C}}$ & $= \frac{\htot{out} - \htot{in}}{u^2}$ & $\checkmark$ & & \\
    Isentropic Efficiency & Losses & $\eisc$ & $= \frac{\htot{out,\, is.} - \htot{in}}{\htot{out} - \htot{in}}$ & $\checkmark$ & $\checkmark$ & \\
    Mass Flow Parameter & Axial Mach Number & $Q_\mathrm{in}$ & $= \frac{\mdot{in}\sqrt{R\ttot{in}}}{\ptot{in}A_{\mathrm{in}}}$ & & $\checkmark$ & $\checkmark$ \\
    Pressure Ratio & Total Pressure Increase & $\prc$ & $ = \frac{\ptot{out}}{\ptot{in}}$ & & $\checkmark$ & $\checkmark$ \\
    Speed Parameter & Circumferential Mach Number & $\hat{N}$ & $= \frac{\pi D N}{\sqrt{R\ttot{in}}}$ & & $\checkmark$ & $\checkmark$ \\
    Torque Parameter & Torque Transmitted by the Shaft & $\hat{M}$ & $= \frac{M}{\mdot{in} D \sqrt{R\ttot{in}}}$ & & & $\checkmark$ \\
    \bottomrule
  \end{tabularx}
  \label{tab:existing-maps}
\end{table*}
Table~\ref{tab:existing-maps} lists the non-dimensional groups used in the three widely used types of compressor performance maps. All three types of maps describe the work exchange between the fluid and the machinery as well as the losses across the component as a function of the inlet conditions. The operation of a single compressor stage can be expressed using the flow coefficient $\varphi$, the work coefficient $\Psi_{\mathrm{C}}$, and the isentropic efficiency $\eisc$~\cite{Howell1950}. By relating the axial flow speed $c_{\mathrm{ax}}$ and the energy input into the fluid $\Delta h_{\mathrm{t}}$ to the blades' rotational speed $u$, the $\varphi$-$\Psi$-$\eta$ maps utilize a single characteristic to describe the performance of a single stage for any shaft speed larger than zero. The isentropic efficiency expresses the losses generated across the  stage by comparing it to an equivalent isentropic compression at the given pressure ratio~\cite{Howell1950}. While modeling multi-stage compressors as a series of compressor stages is possible, overall compressor characteristics are typically preferred~\cite{Howell1950}. The flow coefficient and the work coefficient can be used for such maps in combination with the speed parameter~\cite{Therkorn1992, Riegler1997}. Using inlet mass flow parameter $Q_{\mathrm{in}}$ and pressure ratio $ \prc $ ~\cite{Howell1950,Horlock1958} is a widely used combination. The operating range from locked rotor or wind-milling to start-up and up to maximum power operation has to be covered by today's engine performance models~\cite{Fuksman2012}. However, the flow and work coefficients approach infinity for zero rotational speed and isentropic efficiency features a gap of definition at unity pressure ratio, rendering both the $\varphi$-$\Psi$-$\eta$ and the $Q$-$\Pi$-$\eta$ maps impractical for start-up modeling. Using $Q$-$\Pi$-$\hat{M}$ maps overcomes this gap of definition~\cite{Bretschneider2015}. These express the non-ideal process through the non-dimensional torque supplied at the shaft $\hat{M}$ and the achieved pressure ratio $\prc$~\cite{Riegler2001}. The definitions given in Table~\ref{tab:existing-maps} change in the case of a turbine. The work coefficient, the pressure ratio, and the isentropic efficiencies are defined as
\begin{equation}
  \label{eq:1}
  \Psi_{\mathrm{T}} = \frac{\htot{in} - \htot{out}}{u^2}\mathrm{,}
\end{equation}
\begin{equation}
  \label{eq:2}
  \prt = \frac{\ptot{in}}{\ptot{out}}\mathrm{,}
\end{equation}
and
\begin{equation}
  \label{eq:3}
  \eist = \frac{\htot{in} - \htot{out}}{\htot{in} - \htot{out,\,is.}}
\end{equation}
respectively~\cite{Lewis1996}, and $\hat{M}$ expresses the torque extracted at the shaft. Nevertheless, the discussed limitations remain also in the case of turbine modeling~\cite{Therkorn1994}.

\begin{figure*}[!htb]
  \centering
  \def\svgwidth{0.9\textwidth}
  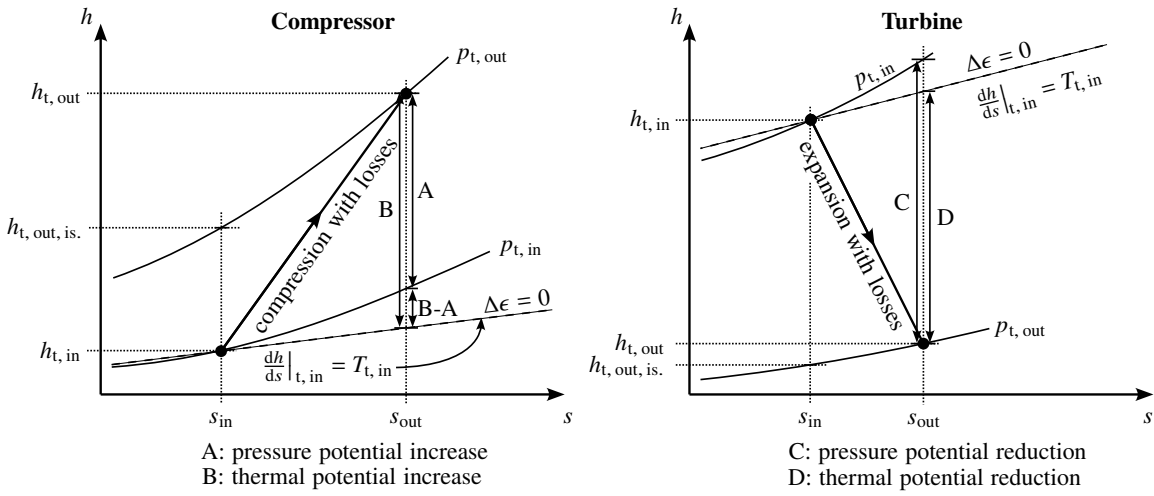
  \caption{Change of state across a turbomachine in an enthalpy-entropy diagram}
  \label{fig:standard-h-s}
\end{figure*}
As shown in Fig.~\ref{fig:standard-h-s}, the total pressure ratio of the compressor describes the increase in pressure potential (A) which can be utilized by expanding from the component exit condition ($\htot{out}, \entr{out}$) to the inlet total pressure ($\ptot{in}$). In the non-isentropic case, further energy can be extracted from the fluid at the compressor outlet by utilizing its remaining thermal potential (B-A) to the constant exergy line ($\Delta \epsilon = 0 $) of the inlet conditions~\cite{zeller2008}. The total pressure ratio of the turbine describes the reduction in pressure potential (C) which is used up by expanding to the component exit condition ($\htot{out}, \entr{out}$) from the inlet total pressure ($\ptot{in}$). In the non-isentropic case, the associated reduction in thermal potential (D) relative to the constant exergy line of the inlet conditions is smaller than the reduction in pressure potential (C). In summary, the change of the fluid's overall thermal potential across a compressor is larger than the pressure ratio indicates whilst the change of the fluid's overall thermal potential across a turbine is smaller than the pressure ratio indicates. A representation of the component characteristics is therefore preferable, which allows tracking the overall thermal potential of the fluid across an entire engine.

The usage of the thermal potential of the fluid takes on a new connotation
in the context of highly integrated transport aircraft with boundary layer ingesting engines. With the conventional methods of thrust-drag bookkeeping becoming more and more involved, analysis of the aircraft as a complete system  is analogous to considering a multi-stage turbomachine as a single system. Power Balance~\cite{Drela2009} and Exergy Analysis~\cite{Arntz2014} provide methods of far-field analysis for this purpose. Considering the design of the engine and its components a part of the aircraft design process~\cite{Riggins1997}, exergy analysis at the engine component level leads to a consistent metric.

In this context, we propose a novel type of turbomachine performance maps based on the change of exergy and anergy across the turbomachine component. Starting from the governing equations, the parameters for the exergy-anergy based performance maps are derived, and their measurability is discussed.  A lossless conversion procedure is introduced between the established and the newly defined types of performance maps. Finally, examples for exergy-anergy based performance maps of a turbine and a compressor are presented.

\section{The Governing Equations of a Turbomachine}

\begin{figure}[!htb]
  \centering
  \def\svgwidth{\linewidth}
  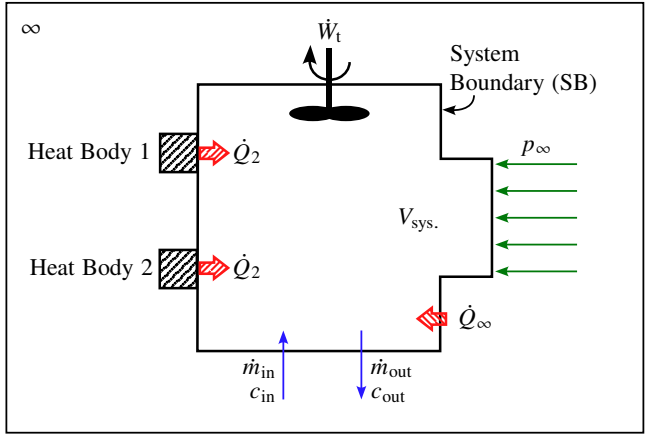
  \caption{A stationary open system in an infinite environment. Adapted from \cite{Weigand2016}.}
  \label{fig:StandardSystem}
\end{figure}

The deduction of the governing equations is based on the principle that turbomachines represent a special case of open systems. Such a generalized open system in an infinitely large environment $\infty$ acting as an energy reservoir is depicted in Fig.~\ref{fig:StandardSystem}.

All fluxes crossing the system boundary into the system are defined as positive, e.g. mass transport ${\dot{m}}$ across the system boundary (SB) is defined positive for mass input into the system. An arbitrary number of heat bodies is shown in Fig.~\ref{fig:StandardSystem}. Each one drives a heat flux $ \dot{Q}_i $ into the system. Mechanical work $\dot{W}_\mathrm{t}$ is exchanged with the system via a shaft. Any pressure differences across the system boundary result in a volume change work ${p}_{\infty}\cdot\frac{dV_\mathrm{sys.}}{dt}$. Any temperature differences across the system boundary result in a heat flux ${\dot{Q}_{\infty}}$ across it. Mass conservation across the system boundary dictates that
\begin{equation}
  \label{eq:4}
    \sum {\left( {\dot{m}} \right)}_{\mathrm{across\,SB}} = \frac{dm_\mathrm{sys.}}{dt} \mathrm{,}
\end{equation}
while momentum conservation becomes
\begin{equation}
  \label{eq:5}
    \sum {\left( {\dot{m} \; c} \right)}_{\mathrm{across\,SB}} + \sum {\left( F \right)}_{\mathrm{across\,SB}} + \sum{F}_{\mathrm{vol.}}= \frac{d \left( m \; c \right)_\mathrm{sys.}}{dt} \mathrm{.}
\end{equation}
Here the $F$ denotes the forces acting on the system boundary. $F_\mathrm{vol.}$ are the volumetric forces. $c$ denotes speed. The energy balance across the system boundary of the open system is given by 
\begin{equation}
  \begin{split}
  \label{eq:6}
    \frac{d}{dt}{\left\{ {U+\dot{m}\;\left(\frac{{c}^{2}}{2}+g \; z\right)} \right\}}_\mathrm{sys.}=\sum {\left[ {{\dot{m}} \; \left(h_{t} + g \; z\right)} \right]}_{\mathrm{across\,SB}}\\
    +\sum {\dot{Q}}_{\mathrm{heat\,flux}}+{\dot{Q}}_{\infty }+{\dot{W}}_{t}-{p}_{\infty } \; \frac{dV_\mathrm{sys.}}{dt} \mathrm{,}
  \end{split}
\end{equation}
where $U$ is the internal energy of the system. $g \; z$ is its potential energy, and $h_{t}$ denotes the total enthalpy. Finally, the entropy transportation equation
\begin{equation}
  \label{eq:7}
  \frac{dS_\mathrm{sys.}}{dt}=\sum (\dot{m} \; s)_{\mathrm{across\,SB}}+\sum (\frac{{\dot{Q}}_{\mathrm{heat\,flux}}}{{T}_{}})+\frac{{\dot{Q}}_{\infty }}{{T}_{\infty }}+({\dot{S}}_\mathrm{prod})_\mathrm{sys.}
\end{equation}
allows considering the second law of thermodynamics for the energy balance, yielding
\begin{equation}
\begin{split}
\label{eq:8}
  -{\dot{W}}_{t}=&-\frac{d}{dt}{\left\{ {U+\dot{m} \; \left( \frac{{c}^{2}}{2}+g \; z \right)+p_{\infty } \; V-T_{\infty } \; s} \right\}}_{\mathrm{sys.}}\\
  & +\sum {\left[ \dot{m} \; \left( {h_{t}}+g \; z-T_{\infty } \; s \right) \right]}_{\mathrm{across\,SB}}\\
  & +\sum {\left( {1-\frac{T_{\infty }}{T_{\mathrm{heat\,flux}}}} \right)} \;{\dot{Q}}_{\mathrm{heat\,flux}} -T_{\infty } \; {\left( {{\dot{S}}_\mathrm{prod}} \right)}_\mathrm{sys.} \mathrm{.}
\end{split}
\end{equation}

\begin{figure}[!htb]
  \centering
  \def\svgwidth{\linewidth}
  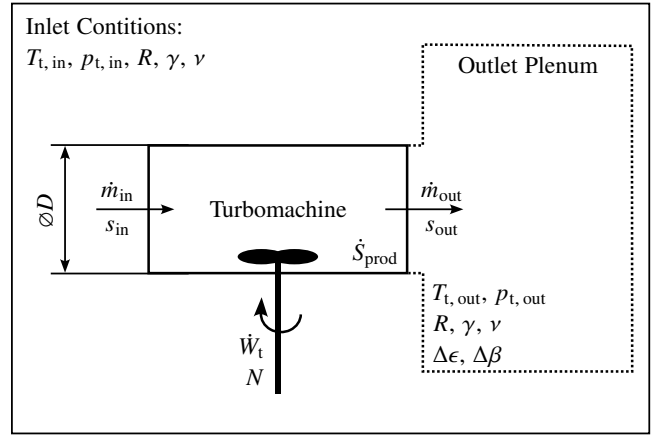
  \caption{Component level performance modelling of a turbomachine}
  \label{fig:turbocomponent-performance}
\end{figure}

In steady-state engine performance computations, the turbomachinery components are assumed to operate between upstream and downstream reservoirs in which kinetic energy is isentropically recovered to result in stagnation properties. This is depicted in Fig.~\ref{fig:turbocomponent-performance}, with the environment representing the upstream reservoir (index in), and the outlet plenum representing the downstream reservoir (index out). The assumption of steady-state operation results in the heat exchange between the fluid and the component structure as well as volume packing in the component to be negligible. Without any bleeds off-takes and leakages, the conservation of mass flow across a turbomachine becomes
\begin{equation}
  \label{eq:9}
  \mdot{in} = \mdot{out} \mathrm{.}
\end{equation}
The equations describing the conservation of mass (Eq.~\eqref{eq:9}) and energy (Eq.~\eqref{eq:6}) are explicitly solved for each component in steady-state performance computations. The momentum equation (Eq.~\eqref{eq:5}) is generally not solved as an independent governing equation. The change of the momentum of the air within the component is realized by the work that is input to or output from the shaft from/to the air along the flow path. This input of work also results in the variation of the gas temperature and pressure that passes through the engine component. Hence, the momentum equation is covered implicitly in the change of the fluid's thermodynamic state.
Consequently, the entropy transport becomes
\begin{equation}
    \label{eq:10}
    \Dot{S}_\mathrm{prod} = \sum \left( \mdot{} \; s \right)_{\mathrm{across\,SB}}
\end{equation}
and the energy balance presented in Eq.~\eqref{eq:8} is simplified to
\begin{equation}
  \label{eq:11}
  -\Dot{W}_\mathrm{t} = \sum \left[\mdot{}\;\left(h_\mathrm{t} - \ttot{in} \; s\right)\right]_{\mathrm{across\,SB}} - \ttot{in} \; \Dot{S}_\mathrm{prod} \mathrm{.}
\end{equation}
Substituting Eq.~\eqref{eq:9} in Eq.~\eqref{eq:11} yields
\begin{equation}
  \label{eq:12}
  \Dot{W}_\mathrm{t} = \mdot{in} \; \left[ \htot{out} - \htot{in} - \ttot{in} \; \left( \entr{out} - \entr{in} \right) \right] + \ttot{in} \; \Dot{S}_\mathrm{prod} \mathrm{.}
\end{equation}
Hence, the work input into the fluid is made up of a reversible part
\begin{equation}
    \label{eq:13}
    \Dot{W}_\mathrm{t,rev} = \mdot{in} \; \left[ \htot{out} - \htot{in} - \ttot{in} \; \left( \entr{out} - \entr{in} \right) \right]
\end{equation}
and an irreversible part
\begin{equation}
    \label{eq:14}
    \Dot{W}_\mathrm{t,irrev} = \ttot{in} \; \Dot{S}_\mathrm{prod}\mathrm{.}
\end{equation}
The reversible part of the work input can be exchanged back between the plenum and the environment. Its mass flow specific form is denoted as the change to the fluid's exergy across the turbomachine, i.e. the outlet's thermal potential to the inlet
\begin{equation}
  \label{eq:15}
  \Delta \epsilon = \htot{out} - \htot{in} - \ttot{in} \; \left( \entr{out} - \entr{in} \right) \mathrm{.}
\end{equation}
Conversely, using the mass specific form of Eq.~\eqref{eq:14} in combination with Eq.~\eqref{eq:10} results in the change to the fluid's anergy
\begin{equation}
  \label{eq:16}
  \Delta \beta = \ttot{in} \; \left( \entr{out} - \entr{in} \right) \mathrm{.}
\end{equation}

\begin{figure*}[!htb]
  \centering
  \def\svgwidth{0.9\textwidth}
  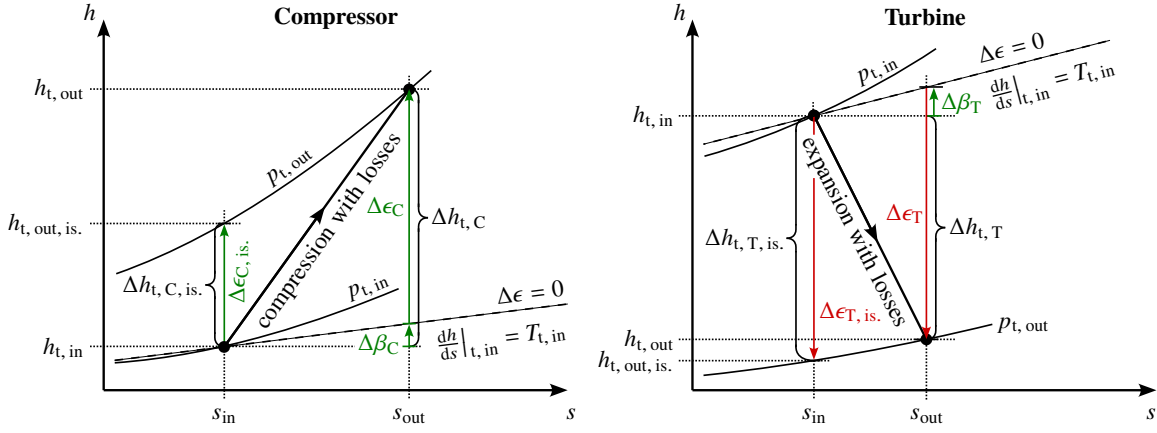
  \caption{The change of exergy and anergy across a turbomachine}
  \label{fig:delta-exan-in-hs}
\end{figure*}
The performance of the turbomachine depicted in Fig.~\ref{fig:turbocomponent-performance} is expressed in terms of the change in the fluid's exergy $ \Delta \epsilon $ and anergy $ \Delta \beta $ caused by the turbomachine. For a machine of characteristic diameter~$D$ this yields
\begin{equation}
  \label{eq:17}
  \Delta \epsilon = f\left( \ttot{in}, \ptot{in}, N, \mdot{in}, D, \gamma, \nu, R \right)
\end{equation}
and
\begin{equation}
  \label{eq:18}
  \Delta \beta = f\left( \ttot{in}, \ptot{in}, N, \mdot{in}, D, \gamma, \nu, R \right) \mathrm{.}
\end{equation}
The properties of the fluid $\gamma$, $R$, and $\nu$ are assumed to be known and available via a numerical model.

\section{The Change of Exergy and Anergy in the Enthalpy-Entropy Diagram}

The change of exergy $\Delta \epsilon$ and anergy $\Delta \beta$ across compressor and turbine have been added to Fig.~\ref{fig:standard-h-s} to give Fig.~\ref{fig:delta-exan-in-hs}. In the case of an ideal, reversible compression process, the energy provided to the fluid via the shaft to achieve a given pressure ratio is $\Delta h_{\mathrm{t,\,C,\,is.}}$ which is equal to $\Delta \epsilon_{\mathrm{C,\,is.}}$. In the case of a real, non-reversible compression process, this amount of energy becomes  $\Delta h_{\mathrm{t,\,C}}$, and Eq.~\eqref{eq:15} and Eq.~\eqref{eq:16} result in
\begin{equation}
  \label{eq:19}
  \Delta h_{\mathrm{t,\,C}} = h_\mathrm{t,\,out} - h_\mathrm{t,\,in} = \Delta \epsilon_\mathrm{C} + \Delta \beta_\mathrm{C} \mathrm{.}
\end{equation}

A positive thermal potential between the component's outlet and its inlet is established, which can be extracted back from the fluid by an arbitrary combination of expansion and cooling to the inlet conditions~\cite{zeller2008}, or via the Carnot cycle. In the real, non-reversible process, however, a part of the total enthalpy increase is irreversibly converted into entropy rather than into thermal potential. This part is revealed by the line of constant exergy ($\Delta \epsilon = 0$). The energy required to reach the outlet entropy level along this line cannot be extracted back from the fluid and is denoted by the change in anergy $\Delta \beta_\mathrm{C}$. In a real, non-reversible compressor, $\Delta \beta_\mathrm{C}$ always has a positive value, while in an ideal compressor $\Delta \beta_{\mathrm{C,\,is.}} = 0$.

Energy is extracted from the fluid and transferred to the shaft in a turbine. In an isentropic expansion process of a given pressure ratio, the drop in exergy is caused solely by the energy extraction from the fluid and thus $\Delta h_{\mathrm{t,\,T,\,is.}} = \Delta \epsilon_{\mathrm{T,\,is.}}$. In the case of a real, non-reversible expansion process of a given pressure ratio, the energy extracted from the fluid becomes $\Delta h_{\mathrm{t,\,T}}$. In this case, Eq.~\eqref{eq:15} and Eq.~\eqref{eq:16} result due to the adopted sign convention in
\begin{equation}
  \label{eq:20}
  \Delta h_{\mathrm{t,\,T}} = h_\mathrm{t,out} - h_\mathrm{t,in} = \Delta \epsilon_\mathrm{T} + \Delta \beta_\mathrm{T} \mathrm{.}
\end{equation}

In a real, non-reversible expansion process, some of the exergy is directly converted into entropy. Again, this part is revealed by the line of constant exergy ($\Delta \epsilon = 0$). In a real, non-reversible turbine, $\Delta \beta_\mathrm{T}$ always has a positive value, while in an ideal expansion process $\Delta \beta_{\mathrm{T,\,is.}} = 0$.

With the inlet conditions known and the properties of the fluid known and available via a numerical fluid model, Eq.~\eqref{eq:19} and Eq.~\eqref{eq:20} yield for compressors and turbines Eq.~\eqref{eq:21}.
\begin{equation}
  \label{eq:21}
  \ttot{out} = f\left( \ttot{in}, \ptot{in}, \Delta \epsilon, \Delta \beta, D, \gamma, \nu, R \right)
\end{equation}
Regarding the working fluid as an ideal gas undergoing a thermodynamic change of state across the turbomachine, the change of entropy becomes~\cite{Weigand2016}
\begin{equation}
  \label{eq:22}
  \Delta s = f\left(\ttot{in}, \ttot{out}, \ptot{in}, \ptot{out}, \gamma, R \right)\mathrm{.}
\end{equation}
According to Eq.~\eqref{eq:16}, it is also attainable as
\begin{equation}
  \label{eq:23}
  \Delta s = \frac{\Delta \beta}{\ttot{in}} = f\left(\Delta \beta,\, \ttot{in} \right)\mathrm{.}
\end{equation}
Combining Eq.~\eqref{eq:21}, Eq.~\eqref{eq:22}, and Eq.~\eqref{eq:23}, and solving for the outlet total pressure, it follows that
\begin{equation}
  \label{eq:24}
  \ptot{out} = f\left( \ttot{in}, \ptot{in}, \Delta \epsilon, \Delta \beta, D, \kappa, \nu, R \right) \mathrm{.}
\end{equation}

Eq.~\eqref{eq:21} and Eq.~\eqref{eq:24} together with Eq.~\eqref{eq:17} and Eq.~\eqref{eq:18} form the basis for the following dimensional analysis.

\section{Dimensional Analysis of the Turbomachine Performance Model}

Equations~\eqref{eq:17} and \eqref{eq:18} provide a complete description of the turbomachine performance using 10 physical parameters, whose units are composed of the fundamental dimensions mass $\unit{\mu}$, length $\unit{\lambda}$, time $\unit{\tau}$, and temperature $\unit{\Theta}$. This is shown in Table~\ref{tab:physicalparam}.
\renewcommand{\arraystretch}{1.3}
\begin{table*}[!htb]
  \caption{Physical parameters employed for turbomachine dimensional analysis}
  \centering
  \begin{tabular}{lccc}
    \toprule
    Name & Symbol & Derived Units & Fundamental Dimensions (SI) \\
    \hline 
    mass flow (inlet) & $\mdot{in}$ & $\unit{\kilogram\; \second^{-1}}$ & $\unit{\mu\; \tau^{-1}}$ \\
    total pressure (inlet) & $\ptot{in}$ & $\unit{\pascal}$ & $\unit{\mu\; \lambda^{-1}\; \tau^{-2}}$\\
    total temperature (inlet) & $\ttot{in}$ & $\unit{\kelvin}$ & $\unit{\Theta}$ \\
    exergy change & $\Delta \epsilon$ & $\unit{\joule\; \kilogram^{-1}}$ &$\unit{\lambda^2\; \tau^{-2}}$ \\
    anergy change & $\Delta \beta$ & $\unit{\joule\; \kilogram^{-1}}$ &$\unit{\lambda^2\; \tau^{-2}}$ \\
    characteristic diameter & $D$ & $\unit{\metre}$ & $\unit{\lambda}$ \\
    shaft speed & $N$ & $\unit{\second^{-1}}$ & $\unit{\tau^{-1}}$ \\
    kinematic viscosity & $\nu$ & $\unit{\metre^2\; \second^{-1}}$ & $\unit{\lambda^2\; \tau^{-1}}$ \\
    ratio of specific heats & $\gamma$ & $\unit{-}$ & $\unit{-}$ \\
    specific gas constant & $R$ & $\unit{\joule\; \kilogram^{-1}\; \kelvin^{-1}}$ & $\unit{\lambda^2}\; \tau^{-2}\; \Theta^{-1}$ \\
    \bottomrule
  \end{tabular}
  \label{tab:physicalparam}
\end{table*}

Combining the parameters into non-dimensional groups reduces the number of system variables and decouples their interdependencies from any reference boundary conditions, e.g. environmental pressure and temperature~\cite{Buckingham1914}. With a total of $P_{\mathrm{init.}} = 10$ initial physical parameters and $P_{\mathrm{rep.}} = 4$ independent repeating variables, a total of $G = (P_{\mathrm{init.}}-P_{\mathrm{rep.}}) = 6$ non-dimensional groups $ \Gamma_i $ result from the dimensional analysis:
\begin{equation}
  \label{eq:25}
  \Gamma_{1} = \frac{\mdot{in}\;\sqrt{R\;\ttot{in}}}{\ptot{in}\; D^{2}} = \mathrm{Q_\mathrm{in}}\mathrm{,}
\end{equation}
\begin{equation}
  \label{eq:26}
  \Gamma_{2} = \frac{N\; D}{\sqrt{R\;\ttot{in}}} = \hat{N}\mathrm{,}
\end{equation}
\begin{equation}
  \label{eq:27}
  \Gamma_{3} = \gamma = \frac{\cp}{c_v}\mathrm{,}
\end{equation}
\begin{equation}
  \label{eq:28}
  \Gamma_{4} = \frac{N\; D^{2}}{\nu} = \mathrm{Re}\mathrm{,}
\end{equation}
\begin{equation}
  \label{eq:29}
  \Gamma_{5} = \frac{\Delta \epsilon}{R\; \ttot{in}} = \epar\mathrm{,}
\end{equation}
and
\begin{equation}
  \label{eq:30}
  \Gamma_{6} = \frac{\Delta \beta}{R\; \ttot{in}} = \bpar\mathrm{.}
\end{equation}

Generally, the proposed exergy-anergy maps are composed of these six non-dimensional groups. If the compressor blades are operated beyond the roughness boundary or the upper critical Reynolds number, the effect of the Reynolds number diminishes \cite{Schaeffler1980}. Otherwise, viscosity effects are 
accounted for via map corrections. Assuming given gas properties, the exergy-anergy maps are represented using four non-dimensional performance parameters: $Q_{\mathrm{in}}$, $\hat{N}$, $\epar$, and $\bpar$. The result is two sets of curves with the curve parameters $\hat{N}$ and $\Delta \epsilon$. These are:
\begin{eqnarray}
  \nonumber
  \frac{\mdot{in}\;\sqrt{R\;\ttot{in}}}{\ptot{in}\; D^{2}} & = & f \left(\epar,\, \frac{N\; D}{\sqrt{R\;\ttot{in}}} \right) \mathrm{,} \\
  \nonumber
  \bpar & = & f \left( \epar,\, \frac{N\; D}{\sqrt{R\;\ttot{in}}} \right) \mathrm{.}
\end{eqnarray}

\section{Experimental Derivation of the Exergy and the Anergy Parameter}

\begin{figure}[!htb]
  \centering
  \def\svgwidth{\linewidth}
  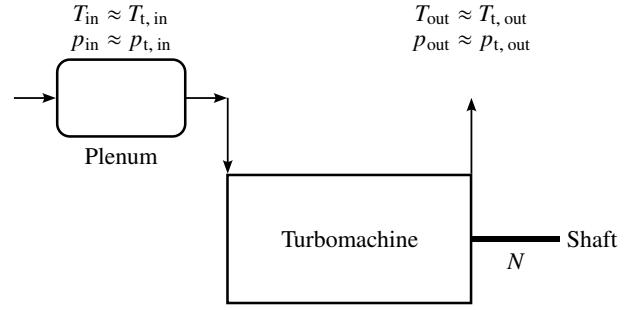
  \caption{Generalized test set-up for a turbomachine}
  \label{fig:test_setup}
\end{figure}

The generalized setup shown in Fig.~\ref{fig:test_setup} is assumed for experimental derivation of the exergy and  the anergy parameter. It applies equally to compressor and turbine testing. The main measurements are entry as well as exit stagnation pressures and temperatures, and the shaft speed.

The exergy parameter
\begin{equation}
  \label{eq:31}
  \epar = \frac{\Delta \epsilon}{R \; \ttot{in}} = \frac{\htot{out} - \htot{in}}{R \; \ttot{in}} - \frac{\entr{out} - \entr{in}}{R}
\end{equation}
and the anergy parameter
\begin{equation}
  \label{eq:32}
  \bpar = \frac{\Delta \beta}{R \; \ttot{in}} = \frac{\entr{out} - \entr{in}}{R}
\end{equation}
shall be derived from the measured parameters shown in Fig.~\ref{fig:test_setup}. For given gas properties,  the first term of Eq.~\eqref{eq:31} is expressed according to~\cite{Weigand2016}:
\begin{equation}
  \begin{split}
    \label{eq:33}
    \frac{\htot{out} - \htot{in}}{R \; \ttot{in}} &= \frac{\cp \; (\ttot{out} - \ttot{in})}{R \; \ttot{in}}\\
    &= \frac{\gamma}{\gamma - 1} \cdot \frac{\ttot{out}- \ttot{in}}{\ttot{in}}\mathrm{.}
  \end{split}
\end{equation}
The change in entropy featured in the second term of Eq.~\eqref{eq:31} and in Eq.~\eqref{eq:32} is a function of the temperature and pressure ratios between the outlet and the inlet~\cite{Weigand2016}, yielding
\begin{equation}
  \label{eq:34}
  \frac{\entr{out} - \entr{in}}{R} = \frac{\gamma}{\gamma - 1} \; \ln \left( \frac{\ttot{out}}{\ttot{in}} \right) - \ln \left( \frac{\ptot{out}}{\ptot{in}} \right) \mathrm{.}
\end{equation}
Both the exergy parameter $\epar$ and the anergy parameter $\bpar$ are functions of the ratio of specific heats of the working fluid, the inlet total pressure $\ptot{in}$, the inlet total temperature $\ttot{in}$, the outlet total pressure $\ptot{out}$, and the outlet total temperature $\ttot{out}$:
\begin{equation}
  \label{eq:35}
  \epar = \frac{\gamma}{\gamma - 1} \; \left( \frac{\ttot{out}}{\ttot{in}} - 1 \right) - \frac{\gamma}{\gamma - 1} \; \ln \left( \frac{\ttot{out}}{\ttot{in}} \right) + \ln \left( \frac{\ptot{out}}{\ptot{in}} \right) \mathrm{,}
\end{equation}

\begin{equation}
  \label{eq:36}
  \bpar = \frac{\gamma}{\gamma - 1} \; \ln \left( \frac{\ttot{out}}{\ttot{in}} \right) - \ln \left( \frac{\ptot{out}}{\ptot{in}} \right) \mathrm{.}
\end{equation}
With given gas properties e.g. available from literature~\cite{Span2019}, the remaining quantities are measured by the typical instrumentation suite of a turbomachine test bed. Consequently, existing testing infrastructure can be utilized for generating the proposed exergy-anergy performance maps.

\section{Converting $Q$-$\Pi$-$\eta$ Maps to Exergy-Anergy-Maps}

Converting $Q$-$\Pi$-$\eta$ maps to exergy-anergy maps requires expressing the exergy parameter and the anergy parameter as functions of the isentropic efficiency $\eta_{\mathrm{is.}}$ and the pressure ratio $\Pi$. As the definitions of the pressure ratio and the isentropic efficiency used in $Q$-$\Pi$-$\eta$ maps differ for compressors and turbines, both types of components must be considered individually.

\subsection{Compressor}
\label{subsec:compressor_to_ea}

In the case of the compressor, the isentropic efficiency is defined as
\begin{equation}
  \label{eq:37}
  \eisc = \frac{\htot{out,\,is.} - \htot{in}}{\htot{out} - \htot{in}}
\end{equation}
Assuming a perfect gas featuring ideal thermal and caloric properties\textcolor{blue}{,} Eq.~\eqref{eq:37} is simplified to
\begin{equation}
  \label{eq:38}
  \eisc = \frac{\ttot{out,\,is.} - \ttot{in}}{\ttot{out} - \ttot{in}}\mathrm{.}
\end{equation}
For a given ratio of specific heats, the ratio of stagnation pressure and temperature of a compressor give
\begin{eqnarray}
  \label{eq:39}
  \frac{\ttot{out}}{\ttot{in}} & = & \frac{\prc^{\frac{\gamma - 1}{\gamma}} - 1 + \eisc}{\eisc} \mathrm{,} \\
  \label{eq:40}
  \prc & = & \frac{\ptot{out}}{\ptot{in}} \mathrm{.}
\end{eqnarray}

This allows to express the exergy parameter of a compressor as a function of the gas properties, the compressor's pressure ratio, and its isentropic efficiency:
\begin{equation}
  \begin{split}
    \label{eq:41}
    \epar = &\;\frac{1}{\eisc} \cdot \frac{\gamma}{\gamma - 1} \; \left( \prc^{\frac{\gamma - 1}{\gamma}} - 1 \right)\\
    &- \frac{\gamma}{\gamma - 1} \; \ln \left( \frac{\prc^{\frac{\gamma - 1}{\gamma}} - 1 + \eisc}{\eisc} \right) + \ln \left( \prc \right) \mathrm{.}
  \end{split}
\end{equation}
Similarly, the anergy parameter of a compressor is expressed in Eq.~\eqref{eq:42} as a function of the gas properties, the compressor's pressure ratio, and its isentropic efficiency.
\begin{equation}
  \label{eq:42}
  \bpar = \frac{\gamma}{\gamma - 1} \; \ln \left( \frac{\prc^{\frac{\gamma - 1}{\gamma}} - 1 + \eisc}{\eisc} \right) - \ln \left( \prc \right)
\end{equation}

\subsection{Turbine}
\label{subsec:turbine_to_ea}

For turbines, the isentropic efficiency is defined as
\begin{equation}
  \label{eq:43}
  \eist = \frac{\htot{in} - \htot{out}}{\htot{in} - \htot{out,\,is.}}\mathrm{.}
\end{equation}
Assuming a perfect gas featuring ideal thermal and caloric properties Eq.~\eqref{eq:43} is simplified to
\begin{equation}
   \label{eq:44} 
   \eist = \frac{\ttot{in} - \ttot{out}}{\ttot{in} - \ttot{out,\,is.}}\mathrm{.}
\end{equation}
For a given ratio of specific heats the ratio of stagnation pressure and temperature of a turbine give 
\begin{eqnarray}
  \label{eq:45}
  \frac{\ttot{out}}{\ttot{in}} & = & 1 - \eist \; \left( 1 - \left( \frac{1}{\prt} \right)^{\frac{\gamma - 1}{\gamma}} \right) \mathrm{,} \\
  \label{eq:46}
  \prt & = & \frac{\ptot{in}}{\ptot{out}}\mathrm{.}
\end{eqnarray}

This allows to express the exergy parameter of the turbine as shown in Eq.~\eqref{eq:47}. It becomes a function of the turbine's pressure ratio and its isentropic efficiency.
\begin{equation}
  \begin{split}
    \label{eq:47}
    \epar = &\;\eist \; \frac{\gamma}{\gamma - 1} \; \left( \prt^{\frac{1 - \gamma}{\gamma}} - 1 \right)\\
            & - \frac{\gamma}{\gamma - 1} \; \ln \left( 1 - \eist \; \left( 1 - \prt^{\frac{1 - \gamma}{\gamma}} \right) \right) - \ln \left( \prt \right)
  \end{split}
\end{equation}
Similarly, the anergy parameter of the turbine is expressed in Eq.~\eqref{eq:48} as a function of the gas properties, the turbine's pressure ratio, and its isentropic efficiency.
\begin{equation}
  \label{eq:48}
  \bpar = \frac{\gamma}{\gamma - 1} \; \ln \left( 1 - \eist \; \left( 1 - \prt^{\frac{1 - \gamma}{\gamma}} \right) \right) + \ln \left( \prt \right) \mathrm{.}
\end{equation}

\section{Converting Exergy-Anergy Maps to $Q$-$\Pi$-$\eta$ Maps}
\label{subsec:backconversion}

The conversion from exergy-anergy maps to $Q$-$\Pi$-$\eta$ maps is different for compressors and turbines due to the differences in the definitions of the isentropic efficiency and the pressure ratio. In the case of the compressor, adding Eq.~\eqref{eq:41} to Eq.~\eqref{eq:42} and solving for the pressure ratio yields
\begin{equation}
  \label{eq:49}
  \prc = \left( \eisc \; \kmoqk \; (\epar + \bpar) + 1 \right)^{\kqkmo} \mathrm{.}
\end{equation}
By substituting $\prc$ back in Eq.~\eqref{eq:42}, the compressor's isentropic efficiency is expressed as
\begin{equation}
  \label{eq:50}
  \eisc = \kqkmo \cdot \frac{e^{-\kmoqk \; \bpar} \; \left( \kmoqk \; \left( \epar + \bpar \right) + 1 \right) - 1}{\epar + \bpar} \mathrm{.}
\end{equation}
It follows from Eq.~\eqref{eq:49} and \eqref{eq:50} that the compressor pressure ratio also becomes a function of $\gamma $, $\epar$, and $\bpar$, namely
\begin{equation}
  \label{eq:51}
  \prc = e^{-\bpar} \; \left( \kmoqk \; \left( \epar + \bpar \right) + 1 \right)^{\kqkmo} \mathrm{.}
\end{equation}
Performing the analogue steps for the turbine results in Eq.~\eqref{eq:52} and Eq.~\eqref{eq:53} for its pressure ratio and isentropic efficiency:
\begin{eqnarray}
  \label{eq:52}
  \prt &=& e^{\bpar} \; \left( \kmoqk \; \left( \epar + \bpar \right) + 1 \right)^{\frac{\gamma}{1 - \gamma}} \mathrm{,} \\
  \label{eq:53}
  \eist &=& \kmoqk \cdot \frac{\epar + \bpar}{e^{-\kmoqk \; \bpar} \; \left( \kmoqk \; \left( \epar + \bpar \right) + 1 \right) - 1} \mathrm{.}
\end{eqnarray}
The isentropic efficiency as well as the pressure ratio of compressors and turbines are attainable in their established form using the exergy parameter, the anergy parameter, and the fluid's ratio of specific heats, enabling conversion between the two map types.

\section{Exergy-Anergy Map Examples}
The form and properties of exergy-anergy maps are discussed using the high-pressure compressor and turbine of a medium bypass ratio, two-spool turbofan engine as an example.

\begin{figure*}[!htb]
  \centering
  \def\svgwidth{1.00\textwidth}
  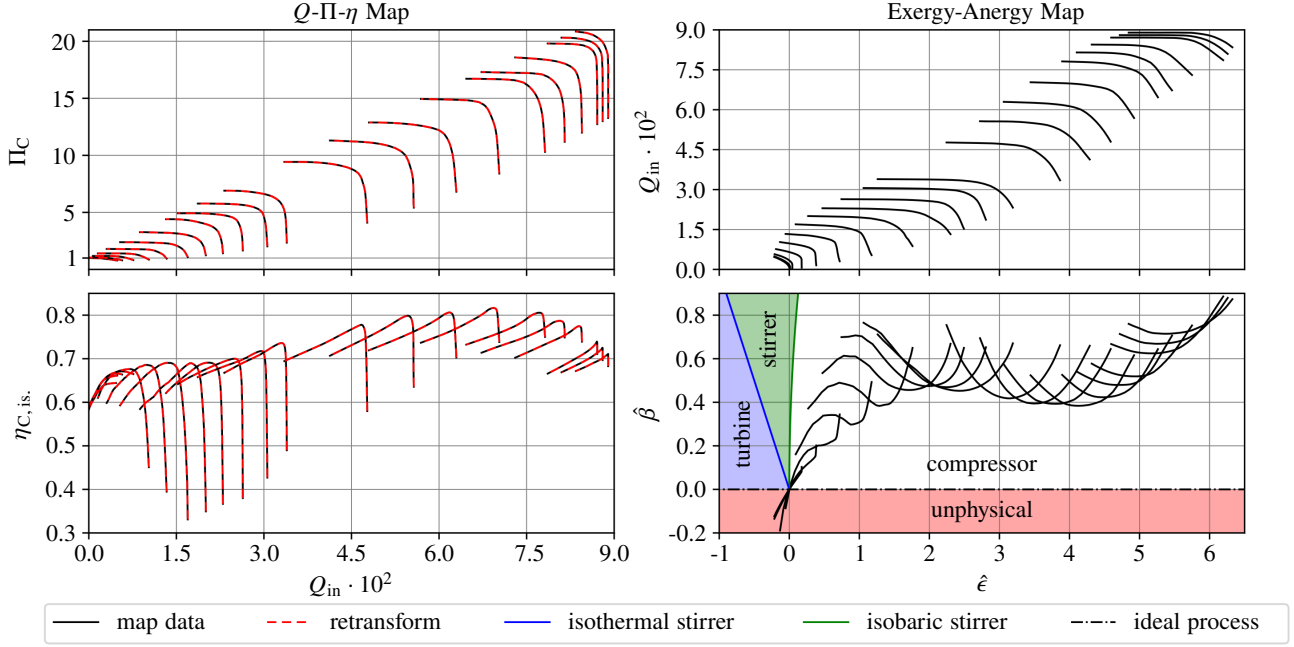
  \caption{Exergy-anergy representation of a compressor map}
  \label{fig:br715-hpc-maps}
\end{figure*}
\subsection{Compressor Map}

The left-hand side of Figure~\ref{fig:br715-hpc-maps} shows the high pressure compressor map in its $Q$-$\Pi$-$\eta$ form with the speed lines depicted as solid black lines. This map has been graphically extrapolated to sub-idle speeds and features data points with a pressure ratio less than unity. The corresponding exergy-anergy representation depicted as solid black lines on the right-hand side of the figure is acquired in accordance with section~\ref{subsec:compressor_to_ea}. The exergy-anergy map data is then converted back to the $Q$-$\Pi$-$\eta$ representation in accordance with section~\ref{subsec:backconversion} and the result is plotted as dashed red lines. The full agreement between this so-called retransform and the $Q$-$\Pi$-$\eta$ map data showcases the complete retention of information during the conversion process.

The $Q_{\mathrm{in}}$-over-$\epar$ plot poses an advantage compared to the representation of $\prc$ over $Q_{\mathrm{in}}$ in the higher speed region, where the speed lines of the exergy-anergy map feature no ranges of constant exergy parameter. This is particularly important in the context of engine performance calculation, where having multiple possible output values for a single input value during map look-up hinders numerical convergence. Therefore, the introduction of a numerical index for each map data point, typical for $Q$-$\Pi$-$\eta$ maps~\cite{Walsh2004}, is not necessary for the calculation of the high-speed operating regimes of high pressure compressors using exergy-anergy maps. However, such numerical indexing is still required for the operating regimes close to and below idle as well as for fans and low pressure compressors.

The $\bpar$-over-$\epar$ plot in the lower right of Fig.~\ref{fig:br715-hpc-maps} shows the change in the exergy and the change in the anergy of the fluid across the compressor. The component's different operating regimes are clearly distinguishable which also allows the assessment of the map's physical consistency. The normal compressor operating regime is marked with a white background. It is characterized by an increase of the fluid's exergy and anergy. It is confined to the downside by the dotted-dashed ideal process line, which describes an isentropic compressor with $\epar > 0$ and $\bpar = 0$. To the left it is confined by the solid green isobaric stirrer line. At the isobaric stirrer line, the energy transferred from the shaft to the fluid is not sufficient to cause a total pressure raise and therefore only an increase in entropy and in total temperature is observed. The energy of the total temperature increase can be extracted back from the fluid, thus it contributes to the increase in thermal potential caused by the compressor. Consequently, $\epar > 0$ along the isobaric stirrer line. If the exergy input from the machinery into the fluid is lower than the dissipation across the compressor, the fluid's thermal potential decreases, although an increase of its total enthalpy is observed. The stirrer operating regime is limited by the solid blue isothermal stirrer line, where the fluid's total enthalpy remains constant, i.e. $\Delta h_{\mathrm{t}} = 0$, and the turbomachine converts all of the fluid's exergy into anergy. If the compressor operates like a turbine it causes a negative change to the fluid's total enthalpy. The part of the ideal process line where $\epar < 0$ describes an isentropic turbine, while the region of the exergy-anergy diagram below the ideal process line is physically not possible, as it violates the second law of thermodynamics. In this use case, the exergy-anergy representation of the compressor's characteristic highlights, that the graphical extrapolation into the sub-idle operating regime is incorrect, because it extends the speed lines into the unphysical region. Such a flaw cannot be noticed in the $Q$-$\Pi$-$\eta$ compressor map, and is prone to lead to inaccurate performance synthesis calculation.

\begin{figure*}[!htb]
  \centering
  \def\svgwidth{1.00\textwidth}
  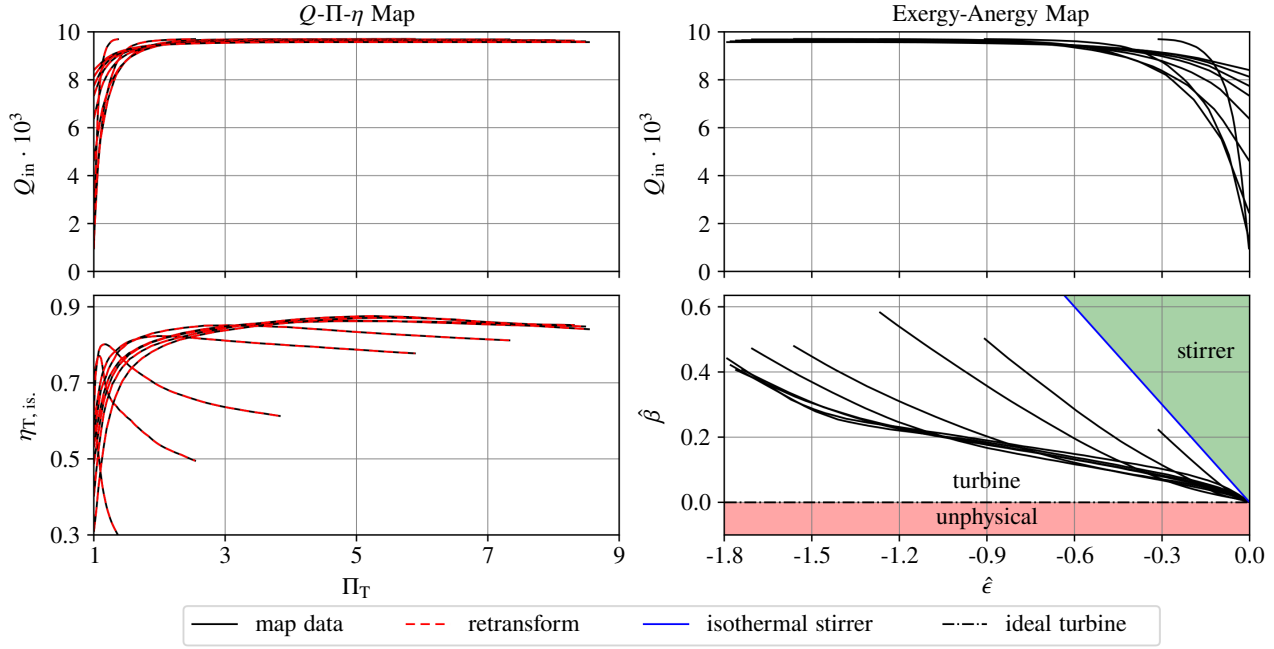
  \caption{Exergy-anergy representation of a turbine map}
  \label{fig:br715-hpt-maps}
\end{figure*}
\subsection{Turbine Map}

The map of the high pressure turbine is depicted in Fig.~\ref{fig:br715-hpt-maps} in both its $Q$-$\Pi$-$\eta$ form and its exergy-anergy representation. The layout is kept the same as in Fig.~\ref{fig:br715-hpc-maps} with the $Q$-$\Pi$-$\eta$ map on the left and the exergy-anergy map on the right-hand side of the figure. The latter was derived from the former by following section~\ref{subsec:turbine_to_ea} and retransformed according to section~\ref{subsec:backconversion}. Again, the retransformed data matches the $Q$-$\Pi$-$\eta$ map data exactly, indicating no information loss during the conversion. The  speed lines in the exergy-anergy map feature no regions of constant exergy parameter, thus no additional numerical indexing is required to ensure the stability of the iterations carried out during performance synthesis calculation.

A decrease of the fluid's exergy, and thus a negative exergy parameter, is observed across a turbine. In the case of Fig.~\ref{fig:br715-hpt-maps}, all map data is within the turbine operating regime marked with a white background in the exergy-anergy map. In this regime, the fluid's exergy is decreased at a higher rate than the rate at which anergy is produced, since energy is transferred from the fluid to the shaft (see Eq.~\eqref{eq:20}). The presented turbine map is deemed physically sound, as no part of it is situated in the unphysical region of the exergy-anergy plot.

\section{Conclusions}

The proposed representation of turbomachinery performance in terms of exergy and anergy leads to a  metric consistent with the exergy-based far-field analysis of future, highly-integrated aircraft. Thermodynamic analysis of turbomachines shows that the change of the fluid's exergy and anergy are sufficient to describe the associated change of the fluid's thermodynamic state.

Dimensional analysis of a generalized turbomachine revealed six non-dimensional groups which are required for a comprehensive description of the thermodynamic process. The non-dimensional groups include the exergy parameter and the anergy parameter. This leads to a novel representation of turbomachine maps. These presented exergy-anergy maps feature no singularities at unity pressure ratio and their definition is universal for compressors and turbines.

It is shown that exergy-anergy maps are obtainable form standard component tests or via the presented lossless conversion of established performance maps. The different operating regimes of turbomachines are clearly distinguishable. Moreover, any physical inconsistencies in the performance maps become visible. It is concluded, that the presented exergy-anergy maps provide an important alternative to the established types of performance maps.

\section*{Acknowledgment} 

The authors gratefully acknowledge funding by the Deutsche Forschungsgemeinschaft (DFG, German Research Foundation). They also would like to thank the colleagues from the Institute for Aerospace Thermodynamics of the University of Stuttgart for their ideas and contributions to this paper.

\section*{Funding Data}

\noindent Funded by the Deutsche Forschungsgemeinschaft (DFG, German Research Foundation) -- Project ID 498601949 -- TRR 364 --.

\begin{nomenclature}
  \entry{$\infty$}{ambient}
  \entry{$c$}{flow speed ($\unit{\metre\: \second^{-1}}$)}
  \entry{$\cp$}{specific heat capacity at constant pressure ($\unit{\joule \: \kilogram^{-1} \: \kelvin^{-1}}$)}
  \entry{$c_v$}{specific heat capacity at constant volume ($\unit{\joule \: \kilogram^{-1} \: \kelvin^{-1}}$)}
  \entry{$D$}{characteristic diameter ($\unit{\metre}$)}
  \entry{$e$}{Euler's number ($\unit{-}$)}
  \entry{$F$}{force ($\unit{\newton}$)}
  \entry{$g$}{gravity acceleration constant ($\SI{9.81}{\metre\:\second^{-2}}$)}
  \entry{$h$}{specific enthalpy ($\unit{\joule \: \kilogram^{-1}}$)}
  \entry{$M$}{torque ($\unit{\newton\:\metre}$)}
  \entry{$m$}{mass ($\unit{\kilogram}$)}
  \entry{$\Dot{m}$}{mass flow ($\unit{\kilogram \: \second^{-1}}$)}
  \entry{$N$}{shaft speed ($\unit{\second^{-1}}$)}
  \entry{$p$}{pressure ($\unit{\pascal}$)}
  \entry{$\Dot{Q}$}{heat flux ($\unit{\joule\: \second^{-1}}$)}
  \entry{$R$}{specific gas constant ($\unit{\joule \: \kilogram^{-1} \: \kelvin^{-1}}$)}
  \entry{$\dot{S}_\mathrm{prod.}$}{entropy production rate ($\unit{\joule \: \kelvin^{-1} \: \second^{-1}}$)}
  \entry{$S$}{entropy ($\unit{\joule \: \kelvin^{-1}}$)}
  \entry{$s$}{specific entropy ($\unit{\joule \: \kilogram^{-1} \: \kelvin^{-1}}$)}
  \entry{$T$}{temperature ($\unit{\kelvin}$)}
  \entry{$t$}{time ($\unit{\second}$)}
  \entry{$U$}{internal energy ($\unit{\joule}$)}
  \entry{$u$}{blades' rotational speed ($\unit{\metre\: \second^{-1}}$)}
  \entry{$V$}{volume ($\unit{\metre^3}$)}
  \entry{$\Dot{W}_{\mathrm{t}}$}{mechanical work ($\unit{J\: \second^{-1}}$)}
  \entry{$z$}{altitude coordinate}

  \EntryHeading{Greek Letters}
  \entry{$\beta$}{specific anergy ($\unit{\joule \: \kilogram^{-1}}$)}
  \entry{$\Delta$}{change}
  \entry{$\epsilon$}{specific exergy ($\unit{\joule \: \kilogram^{-1}}$)}
  \entry{$\Theta$}{temperature, fundamental dimension}
  \entry{$\lambda$}{length, fundamental dimension}
  \entry{$\mu$}{mass, fundamental dimension}
  \entry{$\nu$}{kinematic viscosity ($\unit{\metre^{2}\: \second^{-1}}$)}
  \entry{$\tau$}{time, fundamental dimension}

  \EntryHeading{Non-Dimensional Groups}
  \entry{$\Hat{M}$}{torque parameter, $M / \mdot{in}\:D\:\sqrt{R\:\ttot{in}}$}
  \entry{$\Hat{N}$}{speed parameter, $\pi\:D\:N / \sqrt{R\:\ttot{in}}$}
  \entry{$Q_{\mathrm{in}}$}{mass flow parameter, $\mdot{in}\:\sqrt{R\:\ttot{in}} / \ptot{in}\:A_{\mathrm{in}}$}
  \entry{$\mathrm{Re}$}{Reynolds number, $N\;D^2 / \nu$}
  \entry{$\bpar$}{anergy parameter, $\Delta \beta / R\;\ttot{in}$}
  \entry{$\Gamma$}{non-dimensional group, arbitrary}
  \entry{$\gamma$}{ratio of specific heats, $\cp / c_v$}
  \entry{$\epar$}{exergy parameter, $\Delta \epsilon / R\;\ttot{in}$}
  \entry{$\eisc$}{compressor isentropic efficiency, $(\htot{out,\,is.} - \htot{in}) / (\htot{out} - \htot{in})$}
  \entry{$\eist$}{turbine isentropic efficiency, $(\htot{in} - \htot{out}) / (\htot{in} - \htot{out,\,is.})$}
  \entry{$\Pi_{\mathrm{C}}$}{compressor pressure ratio, $\ptot{out} / \ptot{in}$}
  \entry{$\Pi_{\mathrm{T}}$}{turbine pressure ratio, $\ptot{in} / \ptot{out}$}
  \entry{$\varphi$}{flow coefficient, $c_{\mathrm{ax}} / u$}
  \entry{$\Psi_{\mathrm{C}}$}{compressor work coefficient, $(\htot{out} - \htot{in}) / u^2$}
  \entry{$\Psi_{\mathrm{T}}$}{turbine work coefficient, $(\htot{in} - \htot{out}) / u^2$}

  \EntryHeading{Subscripts}
  \entry{$\infty$}{ambient}
  \entry{$\mathrm{ax}$}{axial}
  \entry{$\mathrm{across\,SB}$}{across the system boundary}
  \entry{$\mathrm{C}$}{compressor}
  \entry{$\mathrm{heat\,flux}$}{heat flux}
  \entry{$\mathrm{in}$}{inlet}
  \entry{$\mathrm{init.}$}{initial}
  \entry{$\mathrm{irrev.}$}{irreversible}
  \entry{$\mathrm{is.}$}{isentropic}
  \entry{$\mathrm{out}$}{outlet}
  \entry{$\mathrm{ref}$}{reference state}
  \entry{$\mathrm{rep.}$}{repeating}
  \entry{$\mathrm{rev.}$}{reversible}
  \entry{$\mathrm{sys.}$}{system}
  \entry{$\mathrm{T}$}{turbine}
  \entry{$\mathrm{t}$}{total value, stagnation value}
  \entry{$\mathrm{vol.}$}{volumetric}

  \EntryHeading{Abbreviations}
  \entry{SB}{system boundary}

\end{nomenclature}    





\nocite{*} 

\bibliographystyle{asmejour}   

\bibliography{references} 



\end{document}